\documentclass[final,authoryear,5p,times]{elsarticle}

\usepackage[utf8]{inputenc}
\usepackage{graphicx}
\usepackage{amssymb}
\biboptions{}

\usepackage{listings}
\newcommand\keywordstyle[1]{\ttfamily #1}
\newcommand\designgoal[1]{\textsl{#1}}
\newcommand\vorent[1]{\textsc{#1}}
\newcommand\rtent[1]{\texttt{\itshape #1}}
\journal{Astronomy \& Computing}

\begin{document}

\begin{frontmatter}



\title{Client Interfaces to the Virtual Observatory Registry}


\author[hd]{Markus Demleitner}
\author[man]{Paul Harrison}
\author[bris]{Mark Taylor}
\author[vopdc]{Jonathan Normand}

\address[hd]{Unversität Heidelberg, Zentrum für Astronomie,
Astronomisches Rechen-Institut, M\"onchhofstraße 12-14, 69120
Heidelberg, Germany}
\address[man]{Jodrell Bank Centre for Astrophysics, Jodrell Bank Observatory, Macclesfield, SK11 9DL, UK}
\address[bris]{H.~H.~Wills Physics Laboratory, Tyndall Avenue, University of Bristol, UK}
\address[vopdc]{Observatoire de Paris VOPDC-IMCCE, 61 Av de l'Observatoire 75014 Paris France}

\begin{abstract}
The Virtual Observatory Registry is a distributed directory of
information systems and other resources relevant to astronomy.  To make
it useful, facilities to query that directory must be provided to humans
and machines alike.  This article reviews the development and status of
such facilities, also
considering the lessons learnt from about a decade of experience 
with Registry interfaces.  
After a brief outline of the history of the standards
development, it describes the use of Registry interfaces in some popular
clients as well as dedicated UIs for interrogating the Registry.  It
continues with a thorough discussion of the design of the two most recent
Registry interface standards, RegTAP on the one hand and a full-text-based
interface on the other hand. The article finally lays out 
some of the less obvious
conventions that emerged in the interaction between providers of
registry records and Registry users as well as
remaining challenges and current developments.

\end{abstract}


\end{frontmatter}

\section{Introduction}
\label{sect:intro}

In \citet{paper1}, henceforth Paper~I, we described the design and
maintenance of the Virtual Observatory (VO) Registry as a distributed
information system.  Conceptually, it is a collection of, by now, about
15000 registry records.  To give the Registry's users --
astronomers, the library
community, or even the general public -- access to this collection,
facilities have to be provided that allow focused queries against it.
This includes common bibliographic constraints (by author,
title or abstract term, year, etc), but also constraints specific to a
registry mainly concerned with
data services (e.g., supported protocols or query parameters,
metadata of published tables).  In the design of such facilities,
several challenges have to be addressed:

\begin{enumerate}
\item different users have very different expectations and requirements
\item the underlying data collection (i.e., the set of registry records)
is changing over time
\item the underlying data structure is fairly complex, and evolves
itself as new standards and techniques are introduced in the VO
\item as many uses require only a small subset of the types of metadata
contained, partial resource descriptions should be retrievable
\item the total data set cannot efficiently be transferred to clients as
a whole
\item registry records are frequently authored by persons not entirely
familiar with the data model, resulting in inconsistent quality
\end{enumerate}

In consequence, no single \emph{user} interface to the Registry can be
sufficient.  Instead, the VO community designed \emph{client}
interfaces,
i.e., network endpoints with rigorously defined behavior and
semantics, designed for use by
programs that then present the actual user interfaces to Registry data.

We will begin this paper with a brief review of the various client
interfaces that are or were used in the VO (section~\ref{sect:history}).
In section~\ref{sect:clients}, we proceed to describe the use some
selected clients make of these facilities and the ways they apply and
expose information obtained from the registry.  A major part of the
paper, section~\ref{sect:regtap}, is devoted to a thorough discussion of
the Registry Relational Model (RegTAP for short), one of the two
registry interfaces currently being developed and deployed in response to the
deficiencies of previous standards.  In section~\ref{sect:fulltext}, the
other new-generation interface is described. 

While laying out some common use cases of Registry data
in section~\ref{sect:commonqueries} we also point out
common query patterns.
Section~\ref{sect:openissues} concludes with some speculation about
probable future developments.

In the following, we refer to common Registry standard texts by their
abbreviated names as introduced in Paper~I, and again the capitalized
word ``Registry'' refers to the abstract concept, while concrete
services are written in lower case (e.g., a ``publishing registry'').
Concepts from VOResource and its extensions are written in
\vorent{small caps}.

\section{History}
\label{sect:history}

Although only explicitly written down in 2011, 
the use cases collected on the IVOA
wiki \citep{wiki:regusecase} outline some of the challenges faced by the
designers of the first client interfaces to the registry in the
mid-2000s -- finding tables containing columns with certain physics,
locating services implementing certain protocols, and the like.

While on the maintenance side of the registry the ecosystem
around OAI-PMH \citep{std:OAIPMH} 
provided guidance for many technology choices,  
in developing the client interfaces much more new ground had to be broken.
For instance, the OPACs (Online Public Access Catalogs; see
\citet{kani2008userperc} for a treatment from about the time of RI1
design)
established in the library community, while comparable for the purpose of
locating information resources,
could not efficiently address the use cases, and no
broadly accepted standard for client, rather than user, interfaces to
OPACs, lent itself to adoption by the VO community.

Given that the interface to be designed was expected to be expressive
enough for requests of the type ``find all TAP services exposing a table
having some word in the description and a column with a given 
UCD\footnote{Unified Content Descriptors or
UCDs in the VO denote phyiscal concepts like ``angular distance'' or
``radio flux'' in a simple formal language \citep{std:UCD}}'',
it was determined fairly early on that an interface based on
simple, atomic parameters would not be sufficent, and
Registry information 
crucial to certain
discovery tasks would not be queryable through it. 
Client interfaces making explicit too much of the
underlying data model would also unduly restrict future developments of
that data model. Thus, at least one
interface to the Registry would have to support a full query
language.  Since the Registry data model was defined in XML Schema, 
an obvious choice for the query language was XQuery
\citep{std:XQUERY}, a language that essentially extends SQL concepts to
querying XML trees.

However, factors against the adoption of XQuery included:
\begin{itemize}
\item the heavy use VOResource makes of XML
namespaces, which tended to make queries hard to write by hand; 
\item the much
larger installed base of relational databases compared to XQuery-capable
engines (compounded by the fact that translating XQuery to a given
relational schema is hard);
\item the desire to open up the full registry data
model to queries written by end users, i.e., astronomers.
As it was expected that many
of these would familiarize themselves with the VO's SQL dialect ADQL
(Astronomical Data Query Language; \citet{std:ADQL}), 
requiring yet another query language
for Registry access appeared undesirable.
\end{itemize}

With these considerations, it was decided to base the primary
Registry interface on conventional relational technology.

While the complex queries XQuery and ADQL allow were needed for
identified use cases, it was also acknowledged that ``Google-like''
searches -- more or less loose matching of words in documents modelled
as bags of words -- was the dominant mode of searching for resources
outside of the VO in the targeted
user base.  At least if common ``comfort'' features like stemming or
phrase searches are desired, this type of search is hard or
impossible to simulate through plain ADQL given its very basic set of
text search capabilities.  Therefore, a keyword search
operation with significant freedom for implementors was also defined.

The result of these considerations was section 2 of RI1
\citep{std:RI1}.
It defines two required search operations
\textit{Search} (with constraints in ADQL) and \textit{KeywordSearch}
(with operator-defined matching of keywords against an
operator-ex\-ten\-sible minimal set of fields) as well as an optional
\textit{XQuerySearch} operation.  All search operations return
either identifier lists or sequences of full resource records in OAI-PMH
style.  In addition, two OAI-PMH-like operations were defined,
\textit{GetResource} to obtain a resource record from an identifier, and
\textit{GetIdentity} to discover metadata about the registry service
itself.

Several implementations of the standards are available; services are
provided by STScI, ESA, and AstroGrid.

As the RI1 design significantly predates the final standardizations of both
ADQL \citep{std:ADQL} and the transport protocol for queries and results
-- that was eventually defined in the TAP standard \citep{std:TAP} --, 
RI1 further defined an ad-hoc transport based on the RPC
mechanism SOAP, and it adopted ADQL at a time when experiments were
underway with passing ADQL statements to client interfaces
in parsed (XML) form.  In
consequence, modern TAP clients cannot use registry endpoints, and
writing queries in the aging XML serialization of ADQL became at least
difficult as software components translating SQL expressions into the
XML forms went unmaintained.

Further critique came from implementor feedback
\citep[e.g.,][]{talk:topcatri1} and was collected together with the use
cases \citep{wiki:regusecase}.  For instance, in practice the
use of a restricted set of XPath to specify constraints instead of
defining an actual relational schema
lead to severe interoperability problems between
different registries, which were further exacerbated by not specifying
rules for case folding.  The apparent flexibility towards registry
extensions provided by the XPath-based column references also did not
pay off as originally expected since registries still needed to do
internal mapping as registry extensions were developed.  In contrast
to the (optional) XQuery interfaces, the (mandatory) ADQL interfaces
frequently lagged behind standards deployment.

In this situation, the most advanced Registry clients relied on the
optional XQuery
interface or even used entirely proprietary interfaces.

As TAP services entered the registry in the early 2010s, RI1's response
format also became a liability.  Registry records contain table
metadata, and with TAP services exposing many tables, resource records 
of several megabytes are not exceptional.  
This made relatively common
queries like ``Retrieve basic metadata on all TAP services'' expensive
in terms of transfer time and processing required.

Therefore, starting in 2011, it was decided to design a new
Registry interface, dubbed ``RESTful'' to contrast it from the RI1
SOAP-based protocol.  
With TAP and ADQL now available, a replacement of the RI1
\textit{Search} operation was mainly a matter of designing a schema and
a mapping to this schema from VOResource. This can be seen as creating a
second serialization of an abstract
data model implicit in VOResource's XML schema files.

The combination of a defined
schema and a TAP service had a model in ObsCore \citep{std:OBSCORE}.
The resulting new standard (``RegTAP''), 
discussed in section~\ref{sect:regtap},
is in the last phases of IVOA peer review as this article
is written.

A replacement for the \textit{KeywordSearch} operation is also being
developed.  Here, the wide availability of feature-rich fulltext engines
such as Apache Lucene
offers the possibility of enriching the bag-of-words model
and allows some advanced operators as well.  We will revisit
this development in section~\ref{sect:fulltext}.

\section{Registry Use in Clients}
\label{sect:clients}

Many VO clients integrate Registry access, frequently without 
advertising the actual source of the data.  Depending on the scope of
the application, different parts of Registry metadata are used, and
different presentations of this information appear appropriate.  In the
following, we look at Registry usage in a number of, we believe,
representative applications, concluding with an in-depth look at
TOPCAT's use of the registry.

\begin{figure}[thm]
\begin{center}
\includegraphics[width=0.35\textwidth]{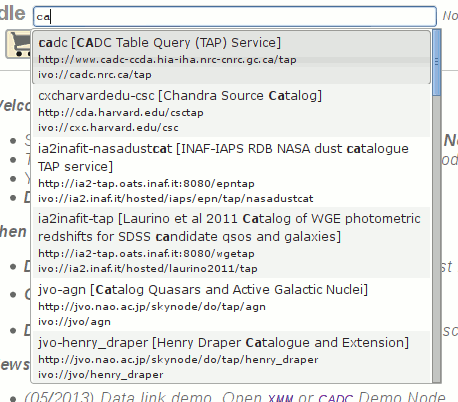}
\end{center}
\caption{TAPHandle uses registry information to provide 
input completion for TAP service access URLs, where the completion items
are complemented by additional metadata.}
\label{fig:TAPHandle}
\end{figure}

TAPHandle\footnote{Online at \texttt{http://saada.unistra.fr/taphandle}.}
is a TAP client operated through web browsers
\citep{2014ASPC..485...15M}.  It uses the Registry
to discover all registered TAP services.  With this information, it can
provide input completion in the selector for the TAP service queried
(Fig.~\ref{fig:TAPHandle}), thus facilitating simple discovery tasks
(``I want to query the CADC TAP server'').  As it is a TAP client, 
is is natural for TAPHandle 
to use RegTAP as its Registry interface. Indeed, its use case is one
of the standard
tasks identified in the collection of requirements for a
revised Registry interface \citep{wiki:regusecase}.
It uses a hard-wired RegTAP endpoint, performing essentially a
single query per session within its
server component. Thus, TAPHandle users are isolated from technical
details of registry access and are also not exposed to visible registry
queries.

Similarly, the spectral analysis tool VOSpec \citep{2005ASPC..347..198O}
queries the registry for all services implementing specific standards
(spectral and line access, in this case),
but since in contrast to TAPHandle
it has no server-side component, it does so directly
from the user's client, using one of two built-in registry endpoints
implementing the RI1 \textit{Search} operation.  Data extraction from
the registry records retrieved is
performed with an
XSLT stylesheet. The discovered resources are
presented to the user in a tree view for individual selection or
de-selection.  This UI is employed both in the selection of the spectral
services and in the selection of servers providing information on the
location of spectral lines.

\begin{figure}[th]
\begin{center}
\includegraphics[width=0.25\textwidth]{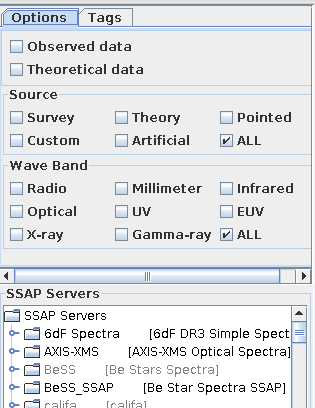}
\end{center}
\caption{SPLAT's rendering of the metadata of spectral services in the
VO: Various metadata obtained from the registry records are made
selectable through checkboxes.}
\label{fig:SPLAT}
\end{figure}

Another VO-enabled spectral tool, SPLAT \citep{2014ascl.soft02008C}, takes this
approach somewhat further by exposing \vorent{dataSource} (from
SimpleDALRegExt, allowing, for instance, the separation of theoretical and
observational services) and \vorent{waveband} from VODataService via
checkboxes in its UI (Fig.~\ref{fig:SPLAT}).  The UI shown is built from
a simple query for all services implementing SSAP. SPLAT furthermore
allows users to add, as it were, private registry records (e.g., for
unpublished services) that are then integrated into this interface.  

A drawback of hiding actual registry queries in this way is that
metadata quality of the resource records directly influences the user
experience for the application itself.  For instance, when a resource
record author neglected to give correct waveband metadata, users knowing
a certain resource serves optical spectra were frequently confused when
the service was deselected after restricting queries to optical data.

The VO client Aladin \citep{soft:Aladin} supporting the major VO
protocols could build upon a registry-like system called GLU that
predates the definition of the VO Registry \citep{2003ASPC..295...43F}.
GLU, with automatic mirror selection and an Aladin-customized
metadata format, to this day distributes Registry
information to Aladin.  Registry records enter GLU not through a client
interface but rather by harvesting an OAI-PMH endpoint.  The operators
of the GLU system at CDS perform additional curation, e.g., by removing invalid
records or records for known-defective services.  By removing all
resource metadata not immediately relevant to the client,
Aladin can keep, in effect, a local cache of the entire
GLU content, which is impractical for
the actual Registry content, as that would currently
entail managing and updating
several hundreds of megabytes.
Responsiveness is further enhanced by persisting this data
between executions of Aladin.

There are also clients specifically built around the Registry.  One of the most
advanced to date is VOExplorer \citep{2008ASPC..394..159T}, developed by
the UK's VO project
AstroGrid in the late 2000s as part of its VODesktop suite.
In its user interface it guides users
in the construction of constraints, mixing menu-based selection with
free-text queries as appropriate.
VOExplorer communicates with registries via the XQuery client interface. 
By thus retrieving only parts of the full VOResource record it
significantly reduces network traffic compared to a RI1 \emph{Search}
client. When a full VOResource record is required, it is cached 
for use in future query results.

Though the major discovery protocols defined while VODesktop was still
being developed are supported, the application's focus is clearly the
integration of Registry data into a workflow, and it offloads
visualization to specialized clients by use of the SAMP
inter-application communication protocol \citep{std:SAMP}.  Regrettably,
VODesktop's development ceased in 2009, with the demise of
the AstroGrid project.

To replace the comprehensive graphical Registry UI provided by
VODesktop,
WIRR\footnote{online at \texttt{http://dc.g-vo.org/WIRR}.} 
was developed.  It is essentially a
browser-based query builder for RegTAP, where, much like in VODesktop,
the user can successively
add constraints on the search results.  Notable constraint types include
queries for
resources containing columns with specific UCDs, ``inverted queries''
to obtain registry information from a service's access URL -- which
is useful for finding contact information when services fail --, or query
with regular expressions on IVORNs.  Even more than VODesktop,
WIRR
relies on external applications to use the resources found, employing
SAMP messages for transmitting resource lists.  TOPCAT is one application
that already supports these.

To support Registry use from within custom user programs, libraries have
been written that encapsulate details of registry access.  Given the
widespread adoption of Astropy \citep{2013A&A...558A..33A}, we note here
the registry functions within the Astropy affiliated
package PyVO \citep{2014ascl.soft02004G}.  For Registry access, it
contains a single function \texttt{regsearch} that supports constraints
by keywords (essentially, a full-text search
within resource record text fields), service types (e.g., image or
spectral service) and wavebands.  The function also has a parameter to
pass in custom SQL fragments executed within the VAO's registry.  Due to
the limitations of RI1 standard client protocols, a custom,
VOTable-based interface is employed at the moment, with a change
to RegTAP in the back-end planned.

Finally, there are uses of Registry client interfaces not directly
connected to actual VO clients.  As an example we mention VO
Fresh\footnote{Online at http://dc.g-vo.org/regrss},
an RSS feed of metadata for services newly
published or updated
in the Registry; new resources are also announced through microblogging
services.  VO Fresh
initially obtained registry information from a full registry's OAI-PMH
endpoint but moved to obtaining registry information through RegTAP as
that became available.

\subsection{Case Study: TOPCAT}

TOPCAT \citep{2005ASPC..347...29T}
is a tool for analysis of astronomical tables.
Part of its function is to provide a user-friendly GUI for
acquiring tabular data from Virtual Observatory services,
most importantly
TAP \citep{std:TAP} and Cone Search \citep{std:SCS},
but also SIA \citep{std:SIAP} and SSA \citep{std:SSAP}.
To achieve this, it needs the Registry to locate
services with the relevant capabilities and to allow the user
to assess their suitability for the science job at hand.

From a user point of view, TOPCAT's registry interaction consists
of selecting a particular type of data service,
optionally supplying some keywords to match against one or
more of a handful of fixed resource metadata fields,
and dispatching a search which results in
presentation of basic metadata for each matching service.
The user then peruses this list and selects one of the returned
services for subsequent use in the application.

\begin{figure}[th]
\begin{center}
\includegraphics[width=0.85\columnwidth]{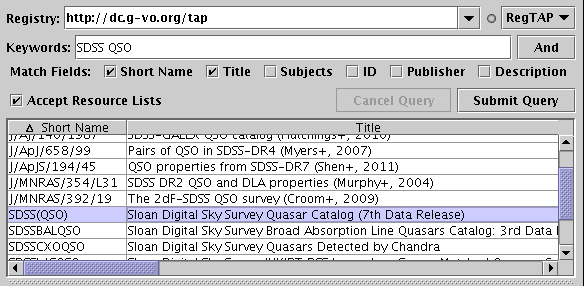}
\end{center}

\caption{TOPCAT's Registry interface: 
the user specifies a query using
a selector for registry service endpoint
and interface protocol (RegTAP or RI1),
a keyword text field,
and a set of checkboxes for what resource fields to match against.
An additional constraint is the service type,
whch depends on the context these widgets are shown in.
Below the input widgets is a listing of matched resources
from which one may be selected.
``Accept Resource List'' allows filling the
resource selector from SAMP messages.}
\label{fig:topcat}
\end{figure}

TOPCAT makes only a single type of registry query to support
this functionality, the user interface to which is illustra\-ted in
Fig.~\ref{fig:topcat}: 
locate all registry resources which offer
a fixed standard capability (e.g.\ TAP) and which
satisfy zero or more additional user search constraints
(e.g.\ ``Title contains the term UKIDSS''),
and for each one return a small fixed amount of metadata
(ID, Title, Publisher, Access URL and a few others).
There is other information stored in the Registry records
that TOPCAT may require, such as {\tt vs:CatalogService}
records describing table and column metadata.
However, for newer VO protocols such information is also available from the
registered data services themselves, and TOPCAT prefers to acquire
it from the latter source, since it may be more reliable and
is also available for unregistered services.

Implementing these queries in RI1 presented some difficulties.
The {\em KeywordSearch\/} operation is unsuitable since keyword
searches cannot be combined with restrictions on service type.
The {\em XQuerySearch\/} operation offers suitable functionality,
but being an optional part of the standard it is only available
from a subset of registry services
(in fact, only the AstroGrid implementation),
and so would
have restricted the choice of registries with which the tool could interact.
The only remaining option is the {\em Search\/} operation.
Syntax and semantics of the fields to match in the required
ADQL queries were somewhat under-documented, and
there is a problem with the way case sensitivity is defined,
but the most serious issue is that {\em Search\/}
always returns the whole, perhaps large, record
for each matched resource, the bulk of which is not needed.
Patchy service implementation quality also contributed to make
RI1-based registry interaction generally slow and unreliable.

With the introduction of RegTAP, 
registry interaction is much improved.
The user interface is almost unchanged, but queries are
more precise,
thanks to more careful mapping of the RM data model
into its relational counterpart,
and much faster,
since it is possible to restrict the query response to
items of interest only.
This latter point can lead to a reduction of two orders of magnitude
in the required data transfer.
To give an admittedly drastic -- but in practice not uncommon --
example, the response size for a query
for all TAP services registered in May 2014
went down to about $150\,\rm kB$ from previously roughly
$25\,\rm MB$.

Note that although for both RI1 and RegTAP the client uses an essentially
SQL-like language to select resources,
what the user sees is a keyword-based or ``Google-like'' interface.
Mapping from the latter to the former can result in verbose
query text, but this text is not difficult for client code to generate.
Therefore, for this purpose there has been no requirement for an
essentially keyword-like client interface to the registry.

There is scope for richer interaction with the registry from TOPCAT,
for instance queries on fine-grained metadata (column UCDs)
or more detailed display of descriptive or curation metadata
from selected records.
These options may be explored in the future.

\section{The Registry Relational Model}
\label{sect:regtap}

\begin{figure*}[t]
\begin{center}
\includegraphics[width=0.75\textwidth]{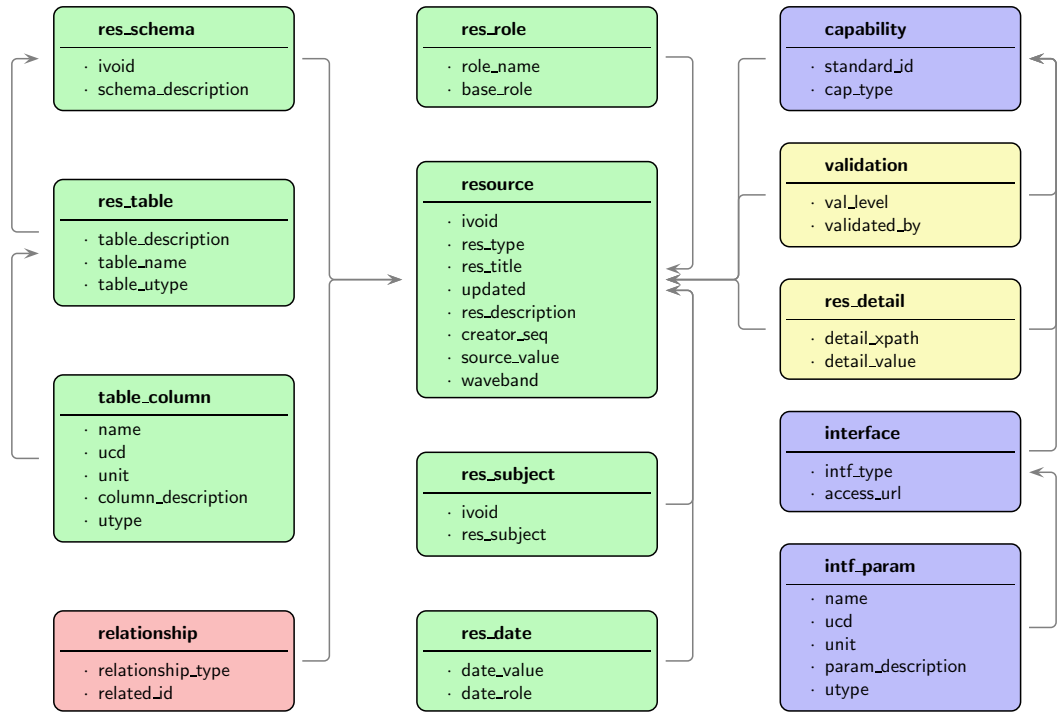}
\end{center}
\caption{A sketch of the database schema of the relational registry,
adapted from \citet{adasspaper}.
The arrows indicate foreign key relationships, the ``attributes''
enumerate the fields most likely interesting to clients or scientists
writing queries.  When joining through
\rtent{relationship} (red) in the discovery of data collection access services
(cf.~section~\ref{sect:relationship}), the green tables would be
``resource-bound'', the blue ones ``capability-bound'', whereas the
yellow ones might be either resource- or capability-bound depending on
query semantics.}
\label{fig:schema}
\end{figure*}

The Registry Relational Model -- briefly called RegTAP for mainly
technical reasons -- is the successor to the \textit{Search} method in
RI1.  It essentially defines a relational schema and rules to map
VOResource records into this schema.  Using TAP as an access
protocol and ADQL as the query language, this is enough to completely
define a client interface to the registry.

A sketch of this relational schema is given in Fig.~\ref{fig:schema}.
Although the authors first experimented with alternative structures that would
have been derived from VOResource algorithmically \citep{wiki:vormap},
it turned out design considerations did not lend themselves to
formalization, as discussed in the next subsection.

\subsection{Design Goals and their consequences}

In the following discussion of RegTAP's design, the model is derived from
several partly conflicting design goals, which are written
\designgoal{slanted} in the following.  Additionally, RegTAP names are 
marked up in
\rtent{slanted typewriter}, while we continue to write
VOResource concepts
in \vorent{small caps}.

While RegTAP attempts to \designgoal{represent all concepts} of
VOResource that could plausibly be of use in locating resources, it is
not a full relational mapping of VOResource.  An overarching design goal
was to \designgoal{keep the model compact}.  In version~1, the model
defines 13 tables, a number that would have been significantly higher
for a full mapping without proportionally adding discovery capabilities.
From VOResource and its current extensions, the model primarily left out:

\begin{itemize}
\item From TAPRegExt \citep{std:TAPREGEXT} the descriptions of user defined 
functions;  these say what extra functions are available in ADQL
queries, how to call them, and what they do.  Representing this would
have required an extra table, which appeared hard to justify given that
no major discovery scenarios were found for this metadata (it is there
for TAP client use, and the TAP clients get the information directly
from the service's capabilities endpoint).
\item Also from TAPRegExt the declaration of how clients can upload
tables into TAP services, with a similar rationale.
\item From StandardsRegExt \citep{std:STDREGEXT} the 
enumerations of the input
parameters defined by the standard itself. They do not appear
valuable for discovery given the moderate number of existing standards.
Representing them would, however, break the simple foreign 
key relationship between interface
and capability if they were kept in the interface table, and an
additional relatively complex table otherwise.
\item Also from StandardRegExt the detailed information on
the versions of documents issued.  This would have required an extra
table, and again, given the moderate number of standards, no credible
discovery scenario is apparent.
\item VOResource's ability to have multiple access URLs for a single
interface; this feature has essentially not been used in practice, and
keeping it would have introduced another join in all queries for access
URLs and hence the vast majority of current Registry queries.
It is planned to drop the feature in future VOResource
versions; resources that actually need multiple access URLs for a single
interface would then have to represent each endpoint as an interface of
its own.
\end{itemize}

To leverage existing VOResource expertise, RegTAP tries to
\designgoal{follow VOResource names}.  However, again compromises had to
be made
to meet some other design goals.  First, as the subject domain of
VOResource partly coincides with the data definition language of SQL,
many of the terms in VOResource are reserved words in database engines.
As RegTAP also tries to \designgoal{avoid requiring delimited
identifiers}\footnote{Delimited identifiers in SQL, syntactically marked
by enclosing the identifier name with double quotes, allow using
arbitrary strings as column names and also suspend SQL's case folding.}
for
usability reasons (e.g., difficult to understand parse errors resulting
from forgotten quotes), conflicting names were amended with tags
indicating entities' roles.  In this way, VOResource's \vorent{table} becomes
\rtent{res\_table}, and \vorent{column} \rtent{ta\-ble\_column}.

Another important design goal was to \designgoal{hide foreign key
relationships}.  This is again a usability concern -- having to
write explicit join conditions would necessitate a more intimate
familiarity with the data model than can be expected from a
possibly casual user.  Instead, query writers should need
only to identify
columns of interest and then use \texttt{NATURAL JOIN} to build their
query's \texttt{FROM} clause.

This implies more name mangling, as in VOResource many elements can be
children of different parents, for instance \vorent{type},
\vorent{name}, \vorent{description}.  Again,
disambiguation is effected
using tags prepended with an underscore, indicating the
source table, abbreviated when names would attain excessive length.
Thus, \vorent{description} in \vorent{resource} becomes
\rtent{res\_description}, whereas in \vorent{capability} it becomes
\rtent{cap\_description}.  Only the two co\-lumn-like tables
(\rtent{table\_column} and \rtent{intf\_param}) are an exception.  
This implies that \rtent{intf\_param} and
\rtent{table\_co\-lumn} are the only tables that cannot be naturally
joined.

The key used for joining is obvious for all tables directly referencing
\rtent{resource}, as Registry semantics ensure \rtent{ivoid} -- the
record's IVORN -- is a
suitable primary key for that table.  However, RegTAP also has foreign
keys into the tables \rtent{capability}, \rtent{interface}, and
\rtent{res\_schema}, for which VOResource does not provide suitable
primary keys, as the respective relationships are represented by lexical
inclusion in XML.
RegTAP instead introduces surrogate keys, the nature of which is
implementation-defined. Hence queries should never explicitly use them,
and since the tables are naturally joinable, they have no reason to do
so.  In general, as the declaration of primary and foreign keys has no
impact on service behavior, RegTAP makes no requirements in this area
but restricts itself to recommendations.

A further design goal requiring changes to VOResource names is that
\designgoal{quoting must not hurt}.  It is not uncommon that SQL authors
and query generators
employ delimited identifiers when they do not need to.
In these cases, mixed-case column names
easily lead to execution errors that again may not be easy to understand.
Therefore, all identifiers in the standard are completely lowercase.  Internal
capitalization to indicate compound words is not uncommon in VOResource,
however.  In RegTAP compound words are concatenated with underscores,
such that, for instance, \vorent{relationshipType} becomes
\rtent{relationship\_type}.

If only for reasons of ease of implementation across different back-end
database engines, it was important for us to \designgoal{not grossly
violate the relational model}.  However, an analogue of the
object-relational impedance mismatch impacts RegTAP as well: for VOResource,
being an XML application, hierarchy and sequences are natural and easy.  In
a relational model, these translate into foreign keys and extra tables
and thus complicate the schema.  In order to avoid an inflation of
tables, RegTAP supports what in effect are arrays of simple strings.

These are only used where 
values are taken from controlled vocabularies, specifically for
\vorent{level} and \vorent{type} from \vorent{content},
\vorent{waveband} and \vorent{rights} from \vorent{Resource},
\vorent{flag} from \vorent{column}, and \vorent{queryType} from
\vorent{interface}.  Here, multiple values from VOResource are
concatenated with hash characters (\#).  To allow reliable querying in
these columns,
RegTAP services must implement an ADQL user defined function called
\texttt{ivo\_hashlist\_has}.  We specifically did not use that pattern
for \vorent{subject}, as its vocabulary, while governed by a
recommendation to use the IVOA Thesaurus, is deliberately open -- indeed, it
is well conceivable that, for instance, hashtags might at some point be
used here -- and it stands to reason that
complex queries over \rtent{res\_subject} will be performed
when clients make use of, say,
ontologies that may themselves be represented in database tables.

In a model so heavily dealing with natural language, another violation
of strict relationality is almost unavoidable: Treating text as, at
least, bags of words.  RegTAP therefore requires conforming services to
offer a user-defined function

\begin{verbatim}
ivo_hasword(txt VARCHAR(*), pat VARCHAR(*)) 
  -> INTEGER
\end{verbatim}

\noindent that returns true at least when \texttt{pat} is present in
\texttt{txt}.  Operators are urged to match \texttt{pat} to \texttt{txt}
in an information-retrieval (IR) sense (i.e., ``Google-style'' as
document vectors).  This is the main violation of RegTAP's design goal that
\designgoal{different registries yield identical results} for identical
queries.  This
violation is regrettable, as experience shows that users are at least
confused if their familiar result lists change after
a change in the registry endpoint used by their client.  However,
given that IR facilities in back-end databases are inconsistent with each
other and an independent implementation of them is nontrivial, the
design goal that the standard \designgoal{does not exclude a 
major database back-end}
overrode the consistency concern.

A final salient design goal is that
\designgoal{Registry extensions are possible without schema updates}.
Registry extensions change the XML schema, and hence RegTAP would have 
to represent arbitrary XML trees within a fixed relational schema if
this design goal were to be fully achieved.  The
result would have been very hard to query indeed.  RegTAP's designers
therefore identified a
subset of extensions that is relatively straightforward
in queries,  powerful enough to satisfy foreseeable use cases, and
reasonably compact: atomic values in 1:n relationships over either
resource or capability. The result is RegTAP's \rtent{res\_detail} table.

This table on the one hand references resources or capabilities by their
\rtent{ivoid} and, as appropriate, the surrogate key on
\rtent{capability} (which is NULL
for items pertaining to the entire resource).   On the other hand it
contains keys (\rtent{detail\_xpath}) and values
(\rtent{detail\_value}).  The keys in this table are essentially XPath
expressions within the resource record, much like the references in RI1
query constraints.  The values are always strings, even when the VOResource
elements represented have other types.  

Thus, a data collection's \vorent{accessURL} child is accessible through
the key \texttt{/accessURL}, the maximum size of files returned from an
image service (defined in SimpleDALRegExt) is retrieved as its decimal
serialization under the key \texttt{/capability/maxFileSize}, and the
authorities managed by a registry are in, if necessary multiple, rows
with the key \texttt{/managedAuthority}.

When mapping existing Registry extensions, it was found this was
sufficient to express the concepts contained with the exceptions
outlined above.  A full list of the keys from the registry extensions
published before RegTAP is given in \citet{std:RegTAP}, and future registry
extensions should specify which additional keys they define.

\subsection{Addressing Particular Issues}

In going from VOResource to a relational schema, properties of either
the relational or the XML model or restrictions of the query language
forced us to introduce additional rules for several entities.  We
mention some major special cases in this subsection.

\paragraph{Case issues} A particular challenge in the mapping rules
from VOResource to RegTAP were case-insensitive values.  For instance,
IVORN \citep{std:VOID}, UCDs, and utypes
in current VO usage \citep{note:utypeusage} all have to be compared
ignoring case.  Even if ADQL had an operator for case-insensitive string
comparison, having to consider case issues in comparisons would
invite bugs in queries that are hard to detect -- when a query author
forgets that a column must be compared ignoring case, the queries
might still return some records and thus appear to work. RegTAP
therefore mandates that all such values must be lowercased during ingestion.  In
this way, queries not taking into account case insensitivity will at
least reliably produce an empty result list.  RegTAP further
case-normalizes other columns filled from controlled vocabularies to
be as consistent as possible.  Only columns intended
for presentation (essentially the descriptions and titles, role name,
and subject) and those where case normalization might lead to
ambiguities (mainly \rtent{detail\_xpath} and \rtent{detail\_value}) are
exempt from normalization. Where case normalized comparisons are desired
for such mixed-case columns, RegTAP offers a UDF \rtent{ivo\_nocase\_match} in
addition to \rtent{ivo\_hasword} (that ignores case as well).

\paragraph{Order} While in most parts of VOResource, the order implied
in XML trees is irrelevant and thus no particular attention is necessary
in the translation to the sets of the relational model, \vorent{creator}
is an exception.  Typically used to convey authorship information, order
there matters to many data providers.  Rather than add sequencing
capabilities to \rtent{res\_role}, RegTAP adds a 
column \rtent{creator\_seq} to \rtent{resource} that contains a
pre-for\-mat\-ted author list.  This has the additional benefit that clients
do not need worry about reconciling the (correct) practice of having one
author per \vorent{creator} element with the (widespread) practice of
including multiple names in one element
in order to produce a flat author list at least for display purposes; 
any necessary special
handling happens at the registry.

\paragraph{QNames} Several VOResource values are really XML qualified
names (QNames).  This concerns some fairly fundamental VOResource
concepts, in particular the types of resources, interfaces, and
capabilities.  For instance, a query might be interested in locating all
resources that are VODataService \vorent{DataCollection}s.  These are
identified by having an XML schema type of
\texttt{\{http://www.ivoa.net/xml/VO\-Data\-Ser\-vice/v1.1\}Da\-taCollection},
using the conventional notation that prepends the namespace part of
a QName in curly brackets.  As this notation is cumbersome for input,
serialized XML maps the namespace URIs to namespace prefixes. 
VOResource and extensions strongly
recommend the use of canonical prefixes that would, for instance, bind
the prefix ``\texttt{vs}'' to
the VODataService namespace.  Hence, the above name becomes a much more
manageable \texttt{vs:Data\-Collection}.  Unfortunately, the canonical
prefixes are not mandatory in VOResource, which means that registries
might use entirely different prefixes, and indeed, in registry practice,
several do.

As long as the RegTAP ingestor knows which attributes contain QNames --
and that is defined in VOResource XML schema files --, it can, however,
unify prefixes by turning the namespace prefixes of the instance
document into namespace URIs and then translating them back into the
canonical prefixes.  To ensure consistent results over registries,
RegTAP requires this prefix normalization.  
Essentially, the recommendation to use its
canonical prefix contained in all VOResource standards becomes a hard
requirement for RegTAP.

\section{Full-text Based Registry Interface}
\label{sect:fulltext}

In parallel with the relatively complex RegTAP interface, a successor to
RI1's \textit{KeywordSearch} is also being developed.
This full-text based registry addresses the difficulty of
extracting information from the previous registry interface. As field
values describing resources are mostly text, a full-text search engine
such as the Apache Lucene library -- as used in popular server components
like ElasticSearch or Apache Solr --
is suitable to index and search
the contents of
the registry. 
It was therefore decided to develop a RESTful
API using ElasticSearch as a client interface to the Registry. The
requirements were to fulfill the registry use cases defined by the IVOA
Registry Working Group \citep{wiki:regusecase} and to support
the web clients developed at VO-Paris Data
Centre, among them several Registry curation tools. 
The full specification of the RESTful interface is currently maintained
at \texttt{http://api.vo.obspm.fr/registry/}. That page also provides
several examples.

\subsection{Query Interface}

The \texttt{/search} method is derived from the \textit{Search} and 
\textit{KeywordSearch}
operations of the RI1 searching interface. It allows querying common
Registry items individually or all together. 
For requests specifying multiple constraints
logical AND is used by default, but a logical OR is available by adding
a parameter ``orValues'' in the query string

The initial set of fields to filter a request has been extended to
fulfill the requirements of all clients using it.  One of the VO client
use cases is to find services by capability type (Spectrum, Image, TAP,
etc),
and specific words or expressions from its
\vorent{description}, \vorent{content.subject}, \vorent{title}, or 
\vorent{shortName} attributes.  Other useful selection criteria come
from \vorent{curation}'s \vorent{publisher}, \vorent{creator.name}, 
and \vorent{contributor}.
The \vorent{coverage} is also used to filter
services in spectral, time, and soon in spatial domain. The spatial
coverage actually is an optional field in the resource description and
not well described in the service declaration. However this information
will soon be available in a standard way, using the HEALPix Multi-Order
Coverage maps (cf.~\ref{sect:coverage}).

As the developers of the full-text search based interfaces are also data
curators in the VO Registry, some specific information was kept and is
available for selected resources:

\begin{itemize}
\item the IVORN of the resource record
\item the dates of the publication and last update of the resource
record
\item the registry where the resource has been declared and which is responsible for this record
\item information from \vorent{validation}.
\end{itemize}

\subsection{Query Examples}

A prototype service implementing the full-text query is being maintained
at VO Paris\footnote{Access URL
\texttt{http://voparis-registry.obspm.fr/vo/ivoa/1/voresources/search}.}.
The following example queries can be executed there by passing them as
URL query strings; for reasons of readability, the query strings are
shown here not URL-encoded.

\begin{itemize}
\item Search all resources containing the keyword ``infrared'':\\
{\small
\hspace*{1em}\texttt{keywords=infrared}
}
 
\item Ditto, but only return services implementing the Simple Image
Access protocol:\\
{\small
\hspace*{1em}\texttt{keywords=infrared}\\
\hspace*{2em}\texttt{+"ivo://ivoa.net/std/SIA"}
}
 
\item  Search for all
resources published by the Centre de données de Strasbourg (CDS)
implementing the Simple Cone Search protocol, with a
\vorent{contentLevel} of Research,
and return the 100 resources starting from match number 200:\\
{\small
\hspace*{1em}\texttt{keywords=publisher:cds}\\
\hspace*{2em}\texttt{+standardid:"ivo://ivoa.net/std/ConeSearch"}\\
\hspace*{2em}\texttt{+contentlevel:Research}\\
\hspace*{1em}\texttt{\&max=100}\\
\hspace*{1em}\texttt{\&from=200}
}
 
\item Return the full resource record for some IVORN:\\
{\small
\hspace*{1em}\texttt{identifier=ivo://vopdc.obspm/luth/exoplanet}
}
 
\end{itemize}

\subsection{Service Response}

As the aim of the full-text query interface is to provide the simplest
system for VO application developers, and most of the new clients are
JavaScript based, the API returns query results formatted in JSON
responses.  Those responses give back the most useful fields as a subset
of the whole service declaration plus a link to the original VOResource
XML file in the registry. The subset of information returned is 
relatively compact and has proven sufficient for the clients already
using the API.

\section{Common Registry Queries}
\label{sect:commonqueries}

VOResource is a complex data model that sometimes
offers multiple ways of expressing apparently very
similar concepts.  Driven by both registry record authoring practices and
the queries employed by popular clients, some usage patterns have evolved that
should be followed for successful Registry use.  Other patterns are
recommended for ease of use. We describe the
patterns in RegTAP terms, but most would be equally
applicable to endpoints speaking XQuery, and partly even to
keyword-based services.

What is special to RegTAP is the query construction technique.  The way
the schema is designed, one looks for the fields to be constrained and
for the fields to be retrieved in a schema description such as
the RegTAP specification, an implementing service's
\texttt{TAP\_SCHEMA}, or its VOSI table metadata.  The query can then be
written by collecting all source tables, concatenating
them by \texttt{NATURAL JOIN} and treating the result as a single table.

\subsection{Locating Standard Services}
\label{sect:capquery}

A very common type of query is finding \vorent{service}s implementing a
certain standard.  In VOResource terms, it is actually not the service
but one of its capabilities that complies with a standard.  For instance,
a service could at the same time implement a cone search for telescope
pointings, and two image services each conforming to a specific version
of SIAP.  Each facility is then represented as a different
\vorent{capability}.

The \rtent{capability} table offers two ways to identify the kind of
interface -- one could constrain \rtent{cap\_type} or
\rtent{standard\_id}.  The correct constraint is on \rtent{standard\_id},
as it would be perfectly legal to register an SSA service, say, with a
\rtent{cap\_type} of \texttt{vr:capability} (i.e., the minimal capability
description only consisting of a standard identifier and the
interfaces).  While the record would miss essential metadata, clients
should have no trouble operating a service registred in this way, and
hence the Registry query should find it.  All known clients' queries by
service type follow the pattern of matching against the standard
identifier.

As the standard identifer is an IVORN, it needs to be lowercased in
queries for RegTAP.  So, to locate all services (say, by their IVORN and
titles) having SSA interfaces, the query would look like this:

\begin{lstlisting}
SELECT ivoid, res_title
FROM rr.resource
  NATURAL JOIN rr.capability
WHERE
  standard_id='ivo://ivoa.net/std/ssa'
\end{lstlisting}

Other relevant standard identifiers are given in the respective
specifications or in one of the examples in RegTAP.

\subsection{Locating Standard Interfaces}
\label{sect:q-standardintf}

Locating the capability is not enough to operate a service. In
addition, the endpoint -- which in VO practice is identified by an
access URL -- needs to be located.  A single capability can have
multiple interfaces, and while this practice is not recommended,
there are resource records that have capabilities
declaring adherence to a standard with interfaces for web browsers in
addition to the standard interface.

VOResource's \vorent{interface} element has a \vorent{role} attribute to
distinguish the standard interfaces from custom ones.  For the former,
\vorent{role} would contain a special string formed according to certain
rules.  In practice, many resource record authors have neglected to set
\vorent{role}, and therefore actual clients started to ignore it.
Current VO practice therefore is to regard the (hopefully unique)
interface of type \vorent{vs:ParamHTTP} as the interface exposing the
standard.

Hence, the pattern to locate interfaces complying to standards right now
is, in RegTAP (this time looking for TAP interfaces):

\begin{lstlisting}
SELECT ivoid, access_url 
FROM rr.capability
  NATURAL JOIN rr.interface
WHERE standard_id='ivo://ivoa.net/std/tap'
  AND intf_type='vs:paramhttp'
\end{lstlisting}

We expect this pattern to be stable, mainly because the development of
StandardsRegExt now very strongly suggests that services supporting
multiple versions of a single standard will have to treat each such
interface in a single capability.  Hence, distinguishing different
versions by the \vorent{role} attribute (or the \rtent{intf\_role}
column in RegTAP) appears dispensable, and assuming the
\vorent{vs:ParamHTTP} interface  within a standard capability must be
the standard service endpoint is straightforward and robust.

\subsection{Query by Physics}

A type of discovery query not yet widely supported in Registry UIs
is the query by
physics.  The fact that resource records can and in many cases do
contain table metadata giving 
UCDs helps locating resources exposing a certain type of data.
As UCDs follow a grammar that ADQL does not understand, it is frequently
advisable to use wildcards in such queries.  For instance,
columns containing infrared magnitudes could be found like this:

\begin{lstlisting}
SELECT name, ucd, column_description
FROM rr.table_column
WHERE ucd LIKE 'phot.mag;em.ir%'
\end{lstlisting}

To illustrate again RegTAP's principle of natural joins, let us
show how to add a constraint on the embedding table here:

\begin{lstlisting}
SELECT name, ucd, column_description,
  table_description
FROM rr.table_column
  NATURAL JOIN rr.res_table
WHERE 1=ivo_hasword(
    table_description, 'quasar')
  AND ucd LIKE 'phot.mag;em.ir%'
\end{lstlisting}

\noindent
-- then constraining on tables accessible via TAP is simply a matter of
combining select list and the FROM and WHERE clauses
from subsection~\ref{sect:q-standardintf} with this query.

\subsection{A Sketch of TOPCAT's Query}

By way of example, we present a query submitted by TOPCAT to
acquire service metadata for presentation to the user,
incorporating some user-supplied constraints.
The following ADQL would locate TAP services concerning galaxies:

\begin{lstlisting}
SELECT ivoid, short_name, res_title, 
  reference_url, base_role, role_name, 
  email, intf_index, access_url, 
  standard_id, cap_type, cap_description, 
  std_version, res_subjects
FROM rr.resource AS res
  NATURAL JOIN rr.interface
  NATURAL JOIN rr.capability
  NATURAL LEFT OUTER JOIN rr.res_role
  NATURAL LEFT OUTER JOIN (
    SELECT 
      ivoid, 
      ivo_string_agg(res_subject, ', ')
        AS res_subjects
     FROM rr.res_subject GROUP BY ivoid
  ) AS sbj
WHERE 
  standard_id='ivo://ivoa.net/std/tap'
  AND intf_type='vs:paramhttp'            
  AND (
    1=ivo_hasword(res_title, 'galaxy')
    OR 1=ivo_hasword(res_subjects, 'galaxy')))
\end{lstlisting}

Note how in this query outer joins are used to make sure rows are
returned even for records that, for instance, do not give roles.  In the
case of \rtent{res\_subject}, VOResource guarantees that at least one
subject must always be present, so doing an outer join here should
not be necessary.  On the other hand, in particular in queries executed
on behalf of a UI, it is good practice to assume minor violations of
VOResource will be present in the Registry.

The sub-query for \rtent{res\_subjects} also shows an example for how to
reduce the number of rows transferred by server-side aggregation.
Another application for this pattern could be, using suitable strings
as separators, retrieving pairs of capability identifiers and their
access URLs.

\section{Open Issues}
\label{sect:openissues}

Even after the introduction of RegTAP, Registry development is not
completed.  In addition to Registry extensions as new service and
resource types are defined within the VO, several fields of work are
currently actively being explored.  We discuss them here as they delineate
what the Registry should be doing but does not do so far, as well as
to document approaches tried in the VO to problems that
may similarly arise in other communities.

\subsection{Data Collection and Relationships}
\label{sect:relationship}

Some TAP services today expose dozens or hundreds of
tables\footnote{Examples for such services include
\texttt{ivo://org.gavo.dc/tap},
\texttt{ivo://nasa.heasarc/services/xamin}, as well as the
TAP interface to VizieR.}.  In ObsCore
services\footnote{The CADC TAP service \texttt{ivo://cadc.nrc.ca/tap}
belongs in this category.} \citep{std:OBSCORE}, data from many
individual data collections are queryable through a single endpoint.  In
the same way, some SIAP services make data from several individual
observatories accessible.

In all these cases, the contributing data collections should all be
present with their full metadata in the Registry. Using GAVO's Lens
Image Archive\footnote{\texttt{ivo://org.gavo.dc/lensunion/q/im}} as an
example, a title query for one of the contributing data collections,
MiNDSTEp, say, should yield the full metadata for the data
collection\footnote{In this case,
\texttt{ivo://org.gavo.dc/danish/red/data}.}, and clients should, from
there, be able to infer the access URL of the service exposing the data.

The \vorent{DataCollection} type of VODataService provides a type for
such cases, and through \vorent{relationship} -- in this case, with a
\vorent{relationshipType} of \texttt{servedBy} -- the associate data
service can be successfully located.

However, client support for querying through \vorent{relationship}
has been lacking, even in the most
advanced registry clients.  WIRR at least shows the presence of
related resources
explicitly, but an additional query is required to retrieve them.
Also, a query for ``Image services exposing data from
MiNDSTEp'' would fail unless the registry record of the embedded service
were carefully crafted.  In RegTAP, writing ADQL for queries that would
simultaneously find ``direct'' (i.e., services exposing exactly one data
collection) and ``indirect'' (i.e., data collection metadata managed
separately from service metadata) services is at least highly nontrivial
\citep{talk:uneasy}.  To understand why joining tables through
relationship requires great care, consider again
Fig.~\ref{fig:schema}.  The colors there distinguish between
``capability-bound'' metadata that in such queries would have to be
queried from the service (e.g., access URL, capability ids,
accepted parameters) and ``resource-bound'' metadata that needs to come
from the data collection itself (e.g., description, title, or UCDs from
a published table).  The two tables that can reference both
\rtent{resource} and \rtent{capability}, shown in yellow in the figure,
additionally complicate query construction.  In any case,
natural joins of tables from different groups will not produce
meaningful results, thus requiring
query authors to add explicit join conditions.

These difficulties have spurred activity to consider changing VOResource
such that clients do not need to follow relationships to locate access URLs.
A reasonable solution explored was adding
(some of) the capabilities of the data service to the resource
records of the data collections themselves.  That, however, leads to an
inflation of such capabilities that will make the very common queries
for all resources implementing a certain standard as laid out in
subsection \ref{sect:capquery} much harder to handle for clients.  For
instance, to prepare for an all-VO query for images, clients would have
filter out duplicate access URLs in order to avoid querying a service
exposing $n$ resources $n$ times (instead of once).

The least burdensome solution is still to be found.  Discussions within the
Registry working group are currently investigating the use of ``auxiliary''
standard identifiers for the capabilities on the data collections,
which would, by lexical convention, facilitate the discovery of either
unique services (by using the standard identifiers already in use) or
all endpoints exposing data constrained by further metadata (by using an
appropriate and index-friendly regular expression).

\subsection{Education and Internationalization}

In the context of work done within the IVOA working group on Education
\citep{note:edumatters}, the issue of multilinguality arose.  While in
professional astronomy, all-English metadata seems sufficient and,
indeed, preferable, the situation is not as clear when certain resources
-- in this case, educational material -- should be made discoverable for
educators or even the general public.  For instance, if a worked-out use-case
on open clusters is available in Italian, should it not be
discoverable by querying for ``Ammasso Aperto''?  This would entail
allowing the relevant text fields (\vorent{title}, \vorent{description},
possibly \vorent{subject}) to be present multiple times in resource
records, each element containing text in a different language, and it
would probably also entail allowing language constraints in client
interfaces to avoid losing precision due to homographs in different
languages.  Alternatively, different registries might be set up for
different languages.

So far, the Education WG only plans to allow discovering which language specific
resources are available in rather than supporting queries in non-English
languages.  If, however, takeup of Registry technologies outside of the
research community were to increase, the issue would have to be
revisited, presumably from both the client and the data model side.

\subsection{Coverage in Space and Time}
\label{sect:coverage}

VODataService allows the specification of resource coverage,
i.e., the spatial area covered on the sky as well as the ranges in time
and spectrum, in resource records.  Apart from the controlled vocabulary
in \vorent{waveband}, this is done through embedding STC-X
\citep{note:STCX} within registry records.  No standard way of querying
this information exists to date.  An attempt to include coverage
information through four tables giving sets of coordinate intervals per
resource as proposed in \citet{talk:ri2} did not gain much traction,
partly because of the flexibility and complexity of the underlying STC
data model, partly because it was felt that for spatial coverage,
coordinate ranges in the equatorial system were too inflexible to be
generally useful even for discovery purposes.  An obvious example
illustrating the shortcomings would be a survey along the galactic
equator: either many ICRS ranges would have to be given, or the coverage
would be dramatically overrepresented.

In the meantime, multi-order coverage maps \citep[][MOCs]{std:MOC}
were developed as a standard way of representing spatial coverages.  Work is
ongoing on how these could be integrated into VOResource on the data
model side and exposed to the clients; if these were to be included in
an extension to RegTAP, 
an ideally indexable way of representing MOCs in databases would
be required, and no technically feasible solution has been proposed so
far.

\section{Conclusions}

The VO Registry is an essential source of metadata about the services
and data that can be used within the VO, and no non-trivial interaction
with the VO can take place without using its discovery capabilities.
Many VO clients embed Registry information and protocols in various
forms.

By necessity, standardization of the Registry protocols occurred
relatively early in the history of the VO.  While standards on the
server side have held out very well, the early standards on the client
side have, in the meantime, proved insufficient for today's advanced
Registry use.

This resulted in the creation of second-generation client interfaces.
In this article, we have discussed principles and design goals of the
two currently developed interfaces.
The keyword search interface provides a simple language to
constrain results that accomodates users' habits in taking up patterns
from general search engines, with a response format designed for easy
integration into browser-based applications.
RegTAP, on the other hand, is a relatively faithful mapping of
essentially the entire data model to a relational database schema,
targeted towards 
``thick'' clients and expert users writing ADQL queries by hand.

While RegTAP goes much further than RI1 in defining the mapping between
the XML schema that defines the Registry data model and the relational
model that is in practice used to represent the data set in queryable
form, there are still some small areas where the relational schema and
its XML counterpart do not precisely match each other's expressiveness,
precluding, for instance, roundtrip ingestion and recreation of registry
records through a RegTAP tableset.  We propose as a lesson to be learned
from this that future data modelling efforts should be done in an
implementation-neutral language with well-defined and well-understood
mappings to the common implementation
languages.  Within the VO, an effort is underway to enable this
\citep{std:VODML}.

This is not to say that the Registry model needs fundamental work or a
technology switch any time soon. The new interfaces to the Registry
expose its functionality fairly completely and interoperably
between their implementations, and building on
proven technologies like TAP and JSON, they also lower the cost of
integrating VO registry information into client programs.

Some open issues remain in the registry's client interface; the most
urgent ones are probably the formulation of contraints on spatial
coverage and the handling of capabilities associated with data
collections.

\section*{Acknowledgements}

We thank Laurent Michel, Pierre Fernique, and Pedro Osuna for providing
information on the Registry interfaces of TAPHandle, Aladin, and VOSpec.

This work was in part supported by the German Astrophysical Virtual
Observatory GAVO, BMBF grant 05A11VH3.

\section*{References}

\bibliographystyle{elsarticle-harv}
\bibliography{regclient}

\begin{thebibliography}{37}
\expandafter\ifx\csname natexlab\endcsname\relax\def\natexlab#1{#1}\fi
\expandafter\ifx\csname url\endcsname\relax
  \def\url#1{\texttt{#1}}\fi
\expandafter\ifx\csname urlprefix\endcsname\relax\def\urlprefix{URL }\fi

\bibitem[{{Astropy Collaboration} et~al.(2013){Astropy Collaboration},
  {Robitaille}, {Tollerud}, {Greenfield}, {Droettboom}, {Bray}, {Aldcroft},
  {Davis}, {Ginsburg}, {Price-Whelan}, {Kerzendorf}, {Conley}, {Crighton},
  {Barbary}, {Muna}, {Ferguson}, {Grollier}, {Parikh}, {Nair}, {Unther},
  {Deil}, {Woillez}, {Conseil}, {Kramer}, {Turner}, {Singer}, {Fox}, {Weaver},
  {Zabalza}, {Edwards}, {Azalee Bostroem}, {Burke}, {Casey}, {Crawford},
  {Dencheva}, {Ely}, {Jenness}, {Labrie}, {Lim}, {Pierfederici}, {Pontzen},
  {Ptak}, {Refsdal}, {Servillat}, and {Streicher}}]{2013A&A...558A..33A}
{Astropy Collaboration}, {Robitaille}, T.~P., {Tollerud}, E.~J., {Greenfield},
  P., {Droettboom}, M., {Bray}, E., {Aldcroft}, T., {Davis}, M., {Ginsburg},
  A., {Price-Whelan}, A.~M., {Kerzendorf}, W.~E., {Conley}, A., {Crighton}, N.,
  {Barbary}, K., {Muna}, D., {Ferguson}, H., {Grollier}, F., {Parikh}, M.~M.,
  {Nair}, P.~H., {Unther}, H.~M., {Deil}, C., {Woillez}, J., {Conseil}, S.,
  {Kramer}, R., {Turner}, J.~E.~H., {Singer}, L., {Fox}, R., {Weaver}, B.~A.,
  {Zabalza}, V., {Edwards}, Z.~I., {Azalee Bostroem}, K., {Burke}, D.~J.,
  {Casey}, A.~R., {Crawford}, S.~M., {Dencheva}, N., {Ely}, J., {Jenness}, T.,
  {Labrie}, K., {Lim}, P.~L., {Pierfederici}, F., {Pontzen}, A., {Ptak}, A.,
  {Refsdal}, B., {Servillat}, M., {Streicher}, O., Oct. 2013. {Astropy: A
  community Python package for astronomy}. A\&A 558, A33.

\bibitem[{Benson et~al.(2009)Benson, Plante, Auden, Graham, Greene, Hill,
  Linde, Morris, O'Mullane, Rixon, Stébé, and Andrews}]{std:RI1}
Benson, K., Plante, R., Auden, E., Graham, M., Greene, G., Hill, M., Linde, T.,
  Morris, D., O'Mullane, W., Rixon, G., Stébé, A., Andrews, K., 2009. {IVOA}
  registry interfaces version 1.0. {IVOA Recommendation}.
\newline\urlprefix\url{http://www.ivoa.net/Documents/RegistryInterface/}

\bibitem[{Boch et~al.(2013)Boch, Donaldson, Durand, Fernique, O'Mullane,
  Reinecke, and Taylor}]{std:MOC}
Boch, T., Donaldson, T., Durand, D., Fernique, P., O'Mullane, W., Reinecke, M.,
  Taylor, M., 2013. {MOC} -- {HEALPix} multi-order coverage map. {IVOA Working
  Draft}.
\newline\urlprefix\url{http://www.ivoa.net/documents/MOC/}

\bibitem[{{Bonnarel} et~al.(2000){Bonnarel}, {Fernique}, {Bienaym{\'e}},
  {Egret}, {Genova}, {Louys}, {Ochsenbein}, {Wenger}, and
  {Bartlett}}]{soft:Aladin}
{Bonnarel}, F., {Fernique}, P., {Bienaym{\'e}}, O., {Egret}, D., {Genova}, F.,
  {Louys}, M., {Ochsenbein}, F., {Wenger}, M., {Bartlett}, J.~G., Apr. 2000.
  {The ALADIN interactive sky atlas. A reference tool for identification of
  astronomical sources}. Astronomy and Astrophysics Supplement 143, 33--40.

\bibitem[{{Castro-Neves} and {Draper}(2014)}]{2014ascl.soft02008C}
{Castro-Neves}, M., {Draper}, P.~W., Feb. 2014. {SPLAT-VO: Spectral Analysis
  Tool for the Virtual Observatory}. Astrophysics Source Code Library.

\bibitem[{Demleitner(2012)}]{talk:ri2}
Demleitner, M., 2012. Towards registry interfaces 2, talk given at the Fall
  2012 IVOA Interop, S\~ao Paulo.
\newline\urlprefix\url{http://docs.g-vo.org/talks/2012-sampa-ri2.pdf}

\bibitem[{Demleitner(2013)}]{talk:uneasy}
Demleitner, M., 2013. News on {RegTAP}, talk given at the Fall 2013 IVOA
  Interop, Waikoloa, HI.
\newline\urlprefix\url{http://wiki.ivoa.net/internal/IVOA/InterOpSep2013Registry/regtap.pdf}

\bibitem[{Demleitner(2014)}]{adasspaper}
Demleitner, M., 2014. {RegTAP} -- a new {API} to the {VO Registry}, poster
  presented at ADASS XIV.

\bibitem[{Demleitner et~al.(2012)Demleitner, Dowler, Plante, Rixon, and
  Taylor}]{std:TAPREGEXT}
Demleitner, M., Dowler, P., Plante, R., Rixon, G., Taylor, M., Aug. 2012.
  {TAPRegExt}: a {VOResource} schema extension for describing {TAP} services,
  version 1.0. {IVOA Recommendation}.
\newline\urlprefix\url{http://www.ivoa.net/Documents/TAPRegExt}

\bibitem[{{Demleitner} et~al.(2014){Demleitner}, {Greene}, {Le Sidaner}, and
  {Plante}}]{paper1}
{Demleitner}, M., {Greene}, G., {Le Sidaner}, P., {Plante}, R.~L., Jul. 2014.
  {The Virtual Observatory Registry}. ArXiv e-prints.

\bibitem[{Demleitner et~al.(2014)Demleitner, Harrison, Molinaro, Greene, Dower,
  and Perdikeas}]{std:RegTAP}
Demleitner, M., Harrison, P., Molinaro, M., Greene, G., Dower, T., Perdikeas,
  M., 2014. {IVOA} registry relational schema. {IVOA Proposed Recommendation}.
\newline\urlprefix\url{http://www.ivoa.net/documents/RegTAP/}

\bibitem[{Derriere et~al.(2004)Derriere, Gray, Mann, Preite~Martinez, McDowell,
  McGlynn, Ochsenbein, Osuna, Rixon, and Williams}]{std:UCD}
Derriere, S., Gray, N., Mann, R., Preite~Martinez, A., McDowell, J., McGlynn,
  T., Ochsenbein, F., Osuna, P., Rixon, G., Williams, R., 2004. {UCD (Unified
  Content Descriptor)} -- moving to {UCD1+}. {IVOA Recommendation}.
\newline\urlprefix\url{http://www.ivoa.net/Documents/latest/UCD.html}

\bibitem[{Dowler et~al.(2010)Dowler, Rixon, and Tody}]{std:TAP}
Dowler, P., Rixon, G., Tody, D., Mar. 2010. Table access protocol version 1.0.
  {IVOA Recommendation}.
\newline\urlprefix\url{http://www.ivoa.net/Documents/TAP}

\bibitem[{{Fernique} et~al.(2003){Fernique}, {Schaaff}, {Bonnarel}, and
  {Boch}}]{2003ASPC..295...43F}
{Fernique}, P., {Schaaff}, A., {Bonnarel}, F., {Boch}, T., 2003. {A Bit of GLUe
  for the VO: Aladin Experience}. In: {Payne}, H.~E., {Jedrzejewski}, R.~I.,
  {Hook}, R.~N. (Eds.), Astronomical Data Analysis Software and Systems XII.
  Vol. 295 of Astronomical Society of the Pacific Conference Series. p.~43.

\bibitem[{Graham et~al.(2013)Graham, Demleitner, Dowler, Fernique, Laurino,
  Lemson, Louys, and Salgado}]{note:utypeusage}
Graham, M., Demleitner, M., Dowler, P., Fernique, P., Laurino, O., Lemson, G.,
  Louys, M., Salgado, J., Feb. 2013. {UTypes}: current usages and practices in
  the {IVOA}. {IVOA Note}.
\newline\urlprefix\url{http://www.ivoa.net/documents/Notes/UTypesUsage}

\bibitem[{{Graham} et~al.(2014){Graham}, {Plante}, {Tody}, and
  {Fitzpatrick}}]{2014ascl.soft02004G}
{Graham}, M., {Plante}, R., {Tody}, D., {Fitzpatrick}, M., Feb. 2014. {PyVO:
  Python access to the Virtual Observatory}. Astrophysics Source Code Library.

\bibitem[{Harrison(2011)}]{wiki:vormap}
Harrison, P., 2011. Relational registry {DM}. {IVOA} Wiki page.
\newline\urlprefix\url{http://wiki.ivoa.net/twiki/bin/view/IVOA/RelationalRegistryDM}

\bibitem[{Harrison et~al.(2012)Harrison, Burke, Plante, Rixon, and
  Morris}]{std:STDREGEXT}
Harrison, P., Burke, D., Plante, R., Rixon, G., Morris, D., May 2012.
  {StandardsRegExt}: a {VOResource} schema extension for describing {IVOA}
  standards, version 1.0. {IVOA Recommendation}.
\newline\urlprefix\url{http://www.ivoa.net/Documents/StandardsRegExt/20120508/REC-StandardsRegExt-1.0-20120508.html}

\bibitem[{{IVOA Registry WG}(2011)}]{wiki:regusecase}
{IVOA Registry WG}, 2011. Requirements and use cases for new registry search
  interface. {IVOA} Wiki page.
\newline\urlprefix\url{http://wiki.ivoa.net/twiki/bin/view/IVOA/RestfulRegistryInterfaceReq}

\bibitem[{Kani-Zabihi et~al.(2008)Kani-Zabihi, Ghinea, and
  Chen}]{kani2008userperc}
Kani-Zabihi, E., Ghinea, G., Chen, S.~Y., 2008. User perceptions of online
  public library catalogues. International Journal of Information Management
  28, 492--502.

\bibitem[{Lemson et~al.(2014)Lemson, Bourg\`es, and Laurino}]{std:VODML}
Lemson, G., Bourg\`es, L., Laurino, O., 9 2014. Vo-dml: a consistent modeling
  language for ivoa data models, {Data Model WG} Internal Working Draft
  2014-09-20.
\newline\urlprefix\url{https://volute.googlecode.com/svn/trunk/projects/dm/vo-dml/doc/VO-DML-WD-v1.0.pdf}

\bibitem[{Louys et~al.(2011)Louys, Bonnarel, Schade, Dowler, Micol, Durand,
  Tody, Michel, Salgado, Chilingarian, Rino, de~Dios~Santander, and
  Skoda}]{std:OBSCORE}
Louys, M., Bonnarel, F., Schade, D., Dowler, P., Micol, A., Durand, D., Tody,
  D., Michel, L., Salgado, J., Chilingarian, I., Rino, B., de~Dios~Santander,
  J., Skoda, P., 2011. Observation data model core components and its
  implementation in the {Table Access Protocol}, version 1.0. {IVOA
  Recommendation}.
\newline\urlprefix\url{http://www.ivoa.net/Documents/ObsCore}

\bibitem[{{Michel} et~al.(2014){Michel}, {Louys}, and
  {Bonnarel}}]{2014ASPC..485...15M}
{Michel}, L., {Louys}, M., {Bonnarel}, F., May 2014. {Browsing TAP Services
  with TAPHandle and DataLink}. In: {Manset}, N., {Forshay}, P. (Eds.),
  Astronomical Society of the Pacific Conference Series. Vol. 485 of
  Astronomical Society of the Pacific Conference Series. p.~15.

\bibitem[{Molinaro et~al.(2014)Molinaro, Demleitner, Ramella, and
  Iafrate}]{note:edumatters}
Molinaro, M., Demleitner, M., Ramella, M., Iafrate, G., 2 2014. Educational
  resources in the virtual observatory, {IVOA} Education {IG} internal note,
  2014-02-25.
\newline\urlprefix\url{http://volute.googlecode.com/svn/trunk/projects/edu/edumatters/edumatters-fmt.html}

\bibitem[{Open Archives Initiative(2002)}]{std:OAIPMH}
Open Archives Initiative, 2002. The open archives initiative protocol for
  metadata harvesting, version 2.0.
\newline\urlprefix\url{http://www.openarchives.org/OAI/openarchivesprotocol.html}

\bibitem[{Ortiz et~al.(2008)Ortiz, Lusted, Dowler, Szalay, Shirasaki,
  Nieto-Santisteba, Ohishi, O'Mullane, Osuna, the VOQL-TEG, and the VOQL
  Working~Group}]{std:ADQL}
Ortiz, I., Lusted, J., Dowler, P., Szalay, A., Shirasaki, Y., Nieto-Santisteba,
  M.~A., Ohishi, M., O'Mullane, W., Osuna, P., the VOQL-TEG, the VOQL
  Working~Group, 2008. {IVOA} astronomical data query language. {IVOA
  Recommendation}.
\newline\urlprefix\url{http://www.ivoa.net/Documents/latest/ADQL.html}

\bibitem[{{Osuna} et~al.(2005){Osuna}, {Barbarisi}, {Salgado}, and
  {Arviset}}]{2005ASPC..347..198O}
{Osuna}, P., {Barbarisi}, I., {Salgado}, J., {Arviset}, C., Dec. 2005. {VOSpec:
  A Tool for Handling Virtual Observatory Compliant Spectra}. In: {Shopbell},
  P., {Britton}, M., {Ebert}, R. (Eds.), Astronomical Data Analysis Software
  and Systems XIV. Vol. 347 of Astronomical Society of the Pacific Conference
  Series. p. 198.

\bibitem[{Plante et~al.(2007)Plante, Linde, Williams, and Noddle}]{std:VOID}
Plante, R., Linde, T., Williams, R., Noddle, K., Mar. 2007. {IVOA} identifiers,
  version 1.03. {IVOA Recommendation}.
\newline\urlprefix\url{http://www.ivoa.net/Documents/REC/Identifiers}

\bibitem[{Robie et~al.(2014)Robie, Chamberlin, Dyck, and Snelson}]{std:XQUERY}
Robie, J., Chamberlin, D., Dyck, M., Snelson, J., 2014. {XQuery} 3.0: An {XML}
  query language. {W3C Recommendation}.
\newline\urlprefix\url{http://www.w3.org/TR/xquery-30/}

\bibitem[{Rots(2005)}]{note:STCX}
Rots, A., Mar. 2005. {STC-X}: Space-time coordinate ({STC}) metadata {XML}
  implementation. {IVOA Note}.
\newline\urlprefix\url{http://www.ivoa.net/documents/latest/STC-X.html}

\bibitem[{Taylor(2010)}]{talk:topcatri1}
Taylor, M., 2010. Registry interfaces 1.0: A service consumer's experience.
  Talk given at the {IVOA} Interop May 2010, Victoria.
\newline\urlprefix\url{http://wiki.ivoa.net/internal/IVOA/InterOpMay2010Reg/reg.pdf}

\bibitem[{Taylor et~al.(2012)Taylor, Boch, Fitzpatrick, Allan, Fay, Paioro,
  Taylor, and Tody}]{std:SAMP}
Taylor, M., Boch, T., Fitzpatrick, M., Allan, A., Fay, J., Paioro, L., Taylor,
  J., Tody, D., 2012. Simple application messaging protocol. {IVOA
  Recommendation}.
\newline\urlprefix\url{http://www.ivoa.net/documents/SAMP/}

\bibitem[{{Taylor}(2005)}]{2005ASPC..347...29T}
{Taylor}, M.~B., Dec. 2005. {TOPCAT {\&} STIL: Starlink Table/VOTable
  Processing Software}. In: {Shopbell}, P., {Britton}, M., {Ebert}, R. (Eds.),
  Astronomical Data Analysis Software and Systems XIV. Vol. 347 of Astronomical
  Society of the Pacific Conference Series. p.~29.

\bibitem[{{Tedds} et~al.(2008){Tedds}, {Winstanley}, {Lawrence}, {Walton},
  {Auden}, and {Dalla}}]{2008ASPC..394..159T}
{Tedds}, J.~A., {Winstanley}, N., {Lawrence}, A., {Walton}, N., {Auden}, E.,
  {Dalla}, S., Aug. 2008. {VOExplorer: Visualising Data Discovery in the
  Virtual Observatory}. In: {Argyle}, R.~W., {Bunclark}, P.~S., {Lewis}, J.~R.
  (Eds.), Astronomical Data Analysis Software and Systems XVII. Vol. 394 of
  Astronomical Society of the Pacific Conference Series. p. 159.

\bibitem[{Tody et~al.(2012)Tody, Dolensky, {McDowell}, Bonnarel, Budavari,
  Busko, Micol, Osuna, Salgado, Skoda, Thompson, and Valdes}]{std:SSAP}
Tody, D., Dolensky, M., {McDowell}, J., Bonnarel, F., Budavari, T., Busko, I.,
  Micol, A., Osuna, P., Salgado, J., Skoda, P., Thompson, R., Valdes, F., 2012.
  Simple spectral access protocol version 1.1. {IVOA Recommendation}.
\newline\urlprefix\url{http://www.ivoa.net/documents/SSA/20120210/REC-SSA-1.1-20120210.htm}

\bibitem[{Tody and Plante(2009)}]{std:SIAP}
Tody, D., Plante, R., 2009. Simple image access specification. {IVOA
  Recommendation}.
\newline\urlprefix\url{http://www.ivoa.net/Documents/latest/SIA.html}

\bibitem[{Williams et~al.(2008)Williams, Hanisch, Szalay, and Plante}]{std:SCS}
Williams, R., Hanisch, R., Szalay, A., Plante, R., 2008. Simple cone search.
  {IVOA Recommendation}.
\newline\urlprefix\url{http://www.ivoa.net/Documents/latest/ConeSearch.html}

\end{thebibliography}

\end{document}